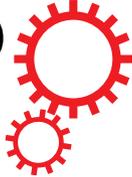

# Transient cerebral hypoperfusion and hypertensive events during atrial fibrillation: a plausible mechanism for cognitive impairment

Matteo Anselmino[1], Stefania Scarsoglio[2], Andrea Saglietto[1], Fiorenzo Gaita[1] & Luca Ridolfi[3]

Atrial fibrillation (AF) is associated with an increased risk of dementia and cognitive decline, independent of strokes. Several mechanisms have been proposed to explain this association, but altered cerebral blood flow dynamics during AF has been poorly investigated: in particular, it is unknown how AF influences hemodynamic parameters of the distal cerebral circulation, at the arteriolar and capillary level. Two coupled lumped-parameter models (systemic and cerebrovascular circulations, respectively) were here used to simulate sinus rhythm (SR) and AF. For each simulation 5000 cardiac cycles were analyzed and cerebral hemodynamic parameters were calculated. With respect to SR, AF triggered a higher variability of the cerebral hemodynamic variables which increases proceeding towards the distal circulation, reaching the maximum extent at the arteriolar and capillary levels. This variability led to critical cerebral hemodynamic events of excessive pressure or reduced blood flow: 303 hypoperfusions occurred at the arteriolar level, while 387 hypertensive events occurred at the capillary level during AF. By contrast, neither hypoperfusions nor hypertensive events occurred during SR. Thus, the impact of AF *per se* on cerebral hemodynamics candidates as a relevant mechanism into the genesis of AF-related cognitive impairment/dementia.

Atrial fibrillation (AF) is the most frequent cardiac arrhythmia and currently affects up to 2% of the general population, with prevalence showing a constant raise due to increased life expectancy[1]. In addition to the well-known AF risk of strokes and TIAs, it has recently emerged that it is also associated with an increased risk of dementia and cognitive decline, independent of clinically relevant events[2].

Several mechanisms have been proposed to explain this association[3–5]: silent cerebral ischemia (SCI), microbleeds, altered cerebral blood flow dynamics and pro-inflammatory conditions are all potential contributors to early cognitive impairment in AF patients. While the role of SCI[6,7] and microbleeds[8,9] in the genesis of cognitive impairment has been investigated, the same is not true for the altered cerebral blood flow. In particular, it is unknown how abnormal heart rhythm influences hemodynamic parameters of the distal cerebral circulation, as determined by time-variable pressure and flow rate signals at the arteriolar and capillary levels.

Given the objective difficulty in measuring those parameters *in vivo*, we simulated AF using two coupled lumped-parameter models of the general cardiovascular[10] and cerebrovascular circulation[11], respectively, and studied cerebrovascular flow dynamics not only in the largest vessels, such as the middle cerebral artery (MCA), but also in the distal cerebral circulation. In our opinion, the results of the present computational approach provide physiopathological insights into the genesis of early dementia and cognitive impairment in AF patients.

[1]Division of Cardiology, Department of Medical Sciences, "Città della Salute e della Scienza" Hospital, University of Turin, Torino, Italy. [2]Department of Mechanical and Aerospace Engineering, Politecnico di Torino, Torino, Italy. [3]Department of Environmental, Land and Infrastructure Engineering, Politecnico di Torino, Torino, Italy. Correspondence and requests for materials should be addressed to S.S. (email: stefania.scarsoglio@polito.it)

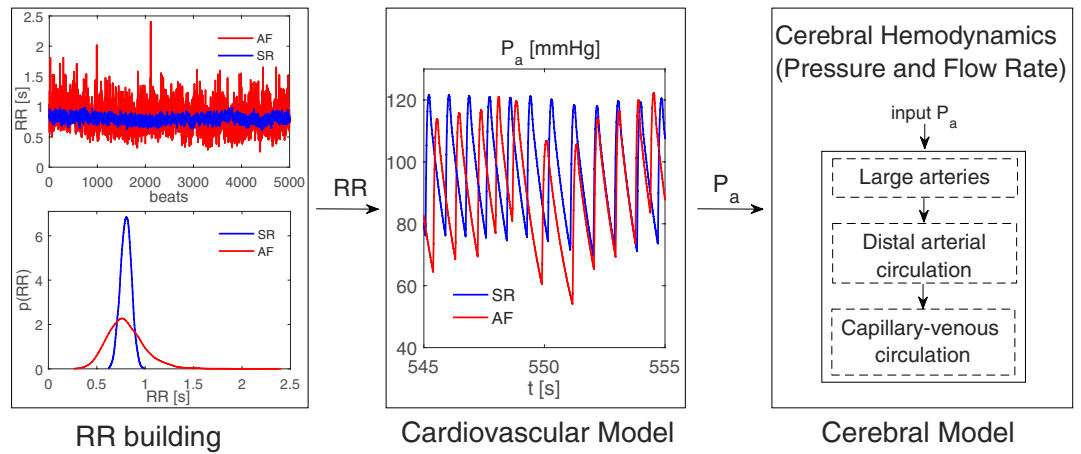

**Figure 1. Schematic representation of the algorithm adopted to model SR and AF intervals in the cerebral dynamics.** Left box: RR series and probability density functions. Central box: representative series for the systemic arterial pressure, $P_a$, as modeled through the cardiovascular model. Right box: cerebral model dynamics, receiving $P_a$ from the cardiovascular model as input. SR: blue curves, AF: red curves.

## Methods

To model sinus rhythm (SR) and AF features in cerebral hemodynamics, we adopted the algorithm outlined in Fig. 1. The first step involves modeling of the RR intervals. SR intervals are extracted from a pink-correlated Gaussian distribution (mean $\mu = 0.8$ [s], standard deviation $\sigma = 0.06$ [s]), as during SR RR intervals are usually Gaussian distributed[12]. The stochastic modeling of RR here adopted refers to sinus rhythm in absence of ectopic beats. AF intervals are instead extracted from an uncorrelated Exponentially Gaussian Modified distribution ($\mu = 0.8$ [s], $\sigma = 0.19$ [s], rate parameter $\gamma = 7.24$ [Hz]), which is unimodal and fully represents the majority of AF cases[12,13]. Both RR series are chosen with the same mean heart rate (75 bpm) to facilitate comparison of the two conditions. In this manner, in fact, AF intervals differ from SR through their uncorrelated nature and higher variability. Five thousand cardiac cycles were simulated in both configurations to guarantee statistical significance. Figure 1 (left box) reports the extracted SR and AF RR series, together with the probability density functions.

Once the RR series were extracted, a lumped parameter model of the cardiovascular system was run to obtain the systemic arterial pressure ($P_a$). The cardiovascular model here used was first proposed by Korakianitis and Shi[14] and has been validated over more than 30 clinical data records regarding AF[10]. It has been also adopted to computationally evaluate the impact of higher HR during AF at rest[15], together with the effect of AF on heart valves[16]. The cardiovascular model includes the systemic and venous circuits for the systemic and pulmonary circulation, together with an active representation of the four cardiac chambers. To mimic AF conditions, both atria are imposed as passive. More details on RR extraction and the cardiovascular model are reported elsewhere[10,16]. Figure 1 (central box) displays two representative $P_a$ time series of SR and AF, as simulated by the cardiovascular model.

The computed $P_a$ was used as the input pressures of the afferent arteries (left and right internal carotid arteries, ICA, and the basilar artery, BA) in the cerebral model (Fig. 1, right box, and section below).

**Cerebral Model.** The present lumped parameter modelling for the cerebrovascular dynamics, first proposed by Ursino and Giannessi[11], is based on the Windkessel approach extended to the whole (arterial and venous) cerebral circulation system. The cerebral model is composed by a network of compliances, C [ml/mmHg], describing the elastic properties of the vasculature, and resistances, R [mmHg s/ml], accounting for the viscous effects. Three cardiovascular variables are involved: the blood flow rate, Q (ml/s), the volume, V (ml), the pressure, P (mmHg). A schematic representation of the cerebral system is displayed in Fig. 2.

The open-loop model can be roughly partitioned into three main sections: large arteries, distal arterial circulation, capillary-venous circulation. The large arteries include the afferent arteries (left and right ICA, and BA) and the circle of Willis (with the left and right precommunicating anterior and posterior cerebral arteries, respectively ACA1 and PCA1, as well as the anterior and posterior communicating arteries, respectively ACoA and PCoA). The six (left and right) main cerebral arteries (middle, MCA, and postcommunicating portions of anterior and posterior cerebral arteries, respectively ACA2 and PCA2) connect the large arteries region to the downstream distal section. The large arteries district is ruled, at each artery node, by a differential equation for mass preservation (accounting for the pressure variation, dP/dt) and pressure-flow rate relations given by the Poiseuille law.

The distal arterial circulation, perfused by the main cerebral arteries, includes the pial circulation and the intracerebral arteries-arterioles, and is divided into six regional districts, independently controlled by autoregulation and $CO_2$ reactivity. The cerebrovascular control mechanisms translate into a temporal variation of the pial arterial-arteriolar compliances, C, and resistances, R. The distal hemodynamics is governed, at each district, by a mass preservation equation (in terms of volume variations, dV/dt), a differential state equation between pressure and volume, and pressure-flow rate relations given by the Poiseuille law. The six microcirculation vascular beds communicate via distal inter-regional anastomoses, represented by resistances.

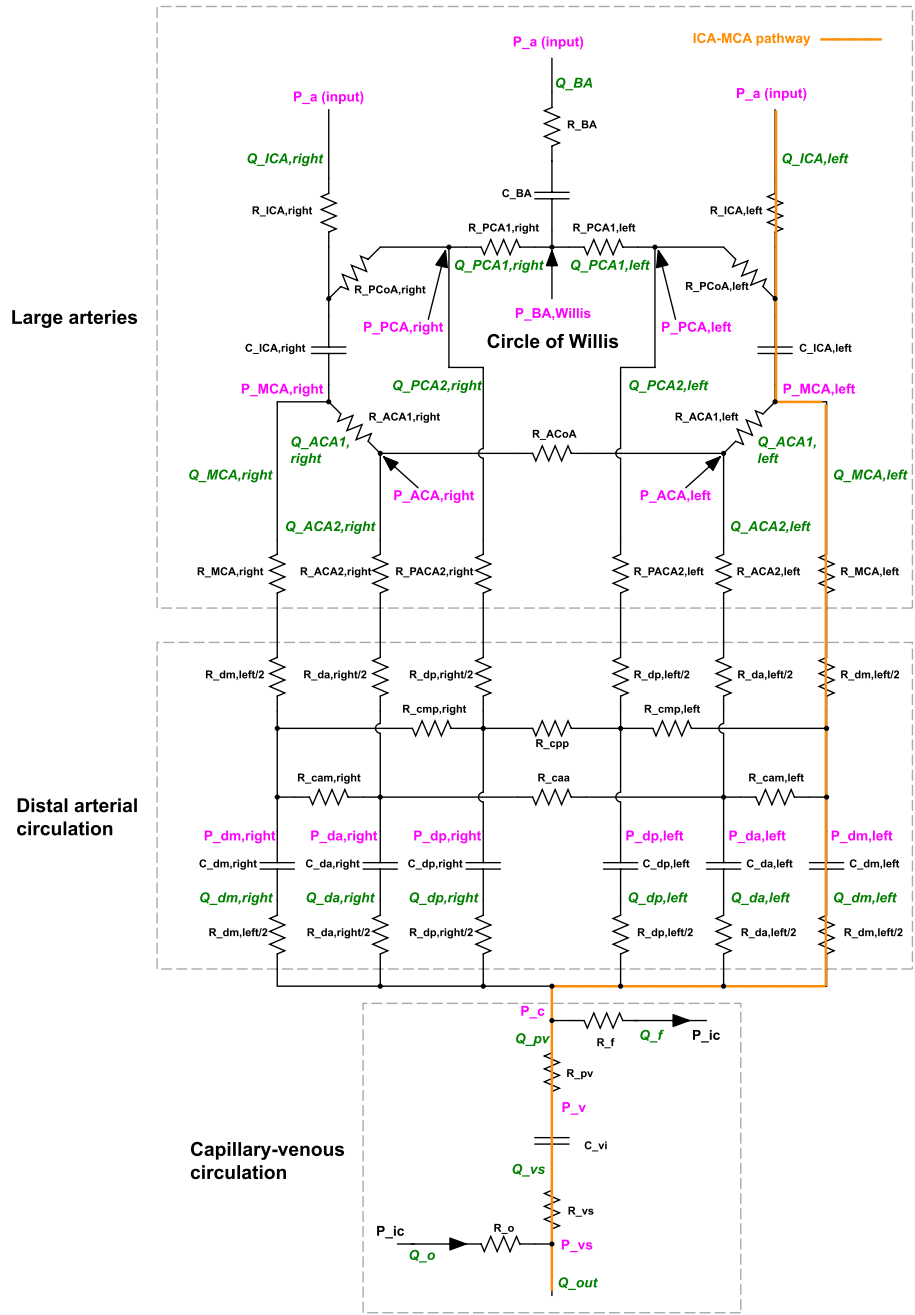

**Figure 2. Scheme of the cerebral model.** R: resistance, C: compliance, Q: flow rate, P: pressure. The left ICA-MCA pathway is highlighted in orange. $P_a$: systemic arterial pressure; $Q_{ICA}$: internal carotid flow rate; $Q_{BA}$: basilar artery flow rate; $P_{BA,willis}$: basilar artery pressure at the entrance of the Circle of Willis; $Q_{ACA1}$: precommunicating anterior cerebral artery flow rate; $Q_{PCA1}$: precommunicating posterior cerebral artery flow rate; $P_{ACA}$: anterior cerebral artery pressure; $P_{MCA}$: middle cerebral artery pressure; $P_{ACA}$: posterior cerebral artery pressure; $Q_{ACA2}$: postcommunicating anterior cerebral artery flow rate; $Q_{PCA2}$: postcommunicating posterior cerebral artery flow rate; $Q_{MCA}$: middle cerebral artery flow rate; $P_{da}$: anterior distal pressure; $P_{dm}$: middle distal pressure; $P_{dp}$: posterior distal pressure; $Q_{da}$: anterior distal flow rate; $Q_{dm}$: middle distal flow rate; $Q_{dp}$: posterior distal flow rate; $P_c$: cerebral capillary pressure; $Q_{pv}$: proximal venous flow rate; $Q_f$: cerebrospinal flow rate at cerebral capillaries; $P_{ic}$: intracranial pressure; $P_v$: cerebral venous pressure; $Q_{vs}$: terminal intracranial venous flow rate; $P_{vs}$: dural sinus pressure; $Q_o$: cerebrospinal flow rate at the dural sinuses; $Q_{out}$: total outflow rate.

Downstream the distal circulation, a unique pressure represents the capillary pressure. The cerebral venous circulation is defined by two-element Windkessel modelling, while the cerebrospinal fluid circulation is formed at the level of cerebral capillaries and is modelled through resistive components. The intracranial pressure value is computed by applying the mass preservation and assuming a monoexponential pressure-volume relationship for the craniospinal system, while pressure-flow rate relations, given by the Poiseuille law, conclude the description.

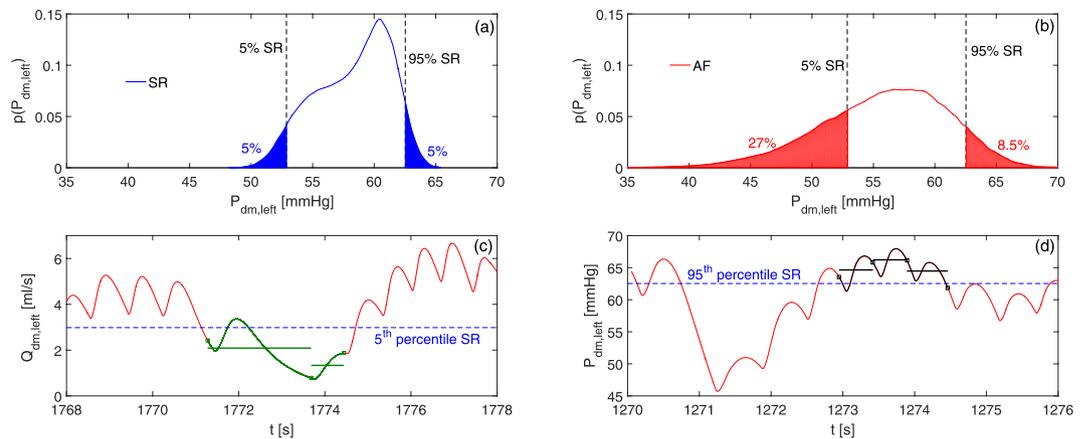

**Figure 3. Examples of percentile, hypoperfusion and hypertensive event evaluation. (a–b)** Example of percentile evaluation in the case of $P_{dm,left}$ ($p(P_{dm,left})$ is the probability density function). SR thresholds (5% SR and 95% SR, dashed lines) individuate the 5$^{th}$ and 95$^{th}$ percentiles in SR (panel a, blue areas), while they correspond to the 27$^{th}$ and 91$^{st}$ percentiles in AF (panel b, red areas). (**c**) Example of hypoperfusion lasting 2 beats ($Q_{dm,left}$). Average flow rate per beats are represented by green horizontal lines, while the 5th percentile SR threshold is displayed through the dashed blue horizontal line. (**d**) Example of hypertensive event lasting 3 beats ($P_{dm,left}$). Average pressure per beats are reported with black horizontal lines, while the dashed blue horizontal line represents the 95th percentile SR threshold.

The model is able to reproduce several different pathological conditions characterized by heterogeneity in cerebrovascular hemodynamics. The complete system of governing differential equations for the cerebral model is reported in the *Supplementary Information*, together with the details of the numerical scheme adopted to solve it (Supplementary Tables S1–S4).

**Data analysis.** The stochastic approach here used allows us to carry out an accurate statistical analysis of the cerebral circulation over 5,000 cardiac cycles. The main statistics (mean and standard deviation) for a number of significant districts are therefore computed and discussed in the following sections.

In order to interpret AF-induced variations, the SR's 5th and 95th percentiles of cerebral hemodynamic variables were taken as "reference thresholds". During AF, we evaluated to which percentile the SR reference thresholds corresponded, thereby quantifying how AF modified the probability of assuming rare values. Figure 3 (panels a and b) provides a graphic example of this evaluation for the probability density function of the middle distal pressure, $p(P_{dm,left})$.

Finally, the prevalence and the duration of cerebral rare events during AF (hypoperfusions and hypertensive events) was evaluated. Hypoperfusion occurs when the average mean flow rate per beat is below the 5th percentile calculated in SR. Hypertension takes place when the average pressure per beat is above the 95th percentile as determined in SR. Both hypoperfusions and hypertensive events can last one single beat or a number of consecutive beats. In Fig. 3c, an example of hypoperfusion lasting 2 beats is reported for $Q_{dm,left}$, while Fig. 3d shows an example of hypertensive event lasting 3 beats for $P_{dm,left}$.

## Results

Mean and standard values of the computed cerebrovascular variables during SR and AF simulations are reported in Table 1.

In particular, among all hemodynamic variables of the cerebral system, we focused on those related to the left ICA-MCA pathway ($P_a$, $Q_{ICA,left}$, $P_{MCA,left}$, $Q_{MCA,left}$, $P_{dm,left}$, $Q_{dm,left}$, $P_c$, $Q_{pv}$), highlighted in orange in Fig. 2, as representative of the blood flow distribution from large arteries to the capillary and venous circulation. Other vascular pathways, namely ICA-ACA ($P_a$, $Q_{ICA,left}$, $P_{MCA,left}$, $Q_{ACA1,left}$, $P_{ACA,left}$, $Q_{ACA2,left}$, $P_{da,left}$, $Q_{da,left}$, $P_c$, $Q_{pv}$) and BA-PCA ($P_a$, $Q_{BA}$, $P_{BA,Willis}$, $Q_{PCA1,left}$, $P_{PCA,left}$, $Q_{PCA2,left}$, $P_{dp,left}$, $Q_{dp,left}$, $P_c$, $Q_{pv}$), present similar values and behaviors.

Figure 4 illustrates representative pressure and flow time series along the left ICA-MCA pathway. Moving from large arteries to the arteriolar-capillary circulation, a progressive increase in the variability of the pressure and flow rate signals occurs. Moreover, in the deep cerebral circulation, AF not only induces more data dispersion than in SR, but also alters the pulsatile dynamics: as a result, increasing/decreasing patterns, lasting up to 5–10 seconds, are observable for the hemodynamic variables of these cerebral regions during AF.

As described in the Methods section, the SR's 5th and 95th percentiles of cerebral hemodynamic variables were taken as reference thresholds to quantify how AF influenced variability of the studied hemodynamic parameters. Figure 5 displays the AF percentiles corresponding to SR 5th and 95th percentiles for variables along the left ICA-MCA pathway (please refer to *Supplementary Information* for the percentiles of the variables along ICA-ACA and BA-PCA pathways: Supplementary Table S5). The distal flow of the MCA region ($Q_{dm,left}$) exhibits a AF value corresponding to the 5th percentile during SR equal to the 15th, accounting for a three-fold increase of probability of a value below that threshold during AF compared to SR simulations. Concomitantly, with regard to cerebral capillary pressure ($P_c$), the AF percentile corresponding to the 95th during SR was the 81th, denoting a nearly four-fold increase of probability to present a value above that threshold (Fig. 5).

| Variable name | Computed values | |
|---|---|---|
| | SR | AF |
| $P_a$ [mmHg] | 99.64 ± 13.95 | 94.77 ± 15.03 |
| $Q_{ICA,left}$ [ml/s] | 4.75 ± 1.73 | 4.71 ± 1.85 |
| $Q_{BA}$ [ml/s] | 3.01 ± 1.13 | 2.99 ± 1.21 |
| $P_{BA,willis}$ [mmHg] | 98.29 ± 13.47 | 93.42 ± 14.57 |
| $Q_{ACA1,left}$ [ml/s] | 1.00 ± 0.39 | 1.00 ± 0.43 |
| $Q_{PCA1,left}$ [ml/s] | 1.50 ± 0.55 | 1.50 ± 0.59 |
| $P_{ACA,left}$ [mmHg] | 93.15 ± 11.60 | 88.31 ± 12.83 |
| $P_{MCA,left}$ [mmHg] | 96.94 ± 13.03 | 92.08 ± 14.11 |
| $P_{PCA,left}$ [mmHg] | 97.14 ± 13.06 | 92.28 ± 14.18 |
| $Q_{ACA2,left}$ [ml/s] | 1.00 ± 0.39 | 1.00 ± 0.43 |
| $Q_{MCA,left}$ [ml/s] | 3.75 ± 1.27 | 3.72 ± 1.39 |
| $Q_{PCA2,left}$ [ml/s] | 1.50 ± 0.55 | 1.49 ± 0.59 |
| $P_{da,left}$ [mmHg] | 58.28 ± 4.29 | 55.84 ± 7.53 |
| $P_{dm,left}$ [mmHg] | 58.28 ± 3.05 | 55.82 ± 5.28 |
| $P_{dp,left}$ [mmHg] | 58.37 ± 3.65 | 55.92 ± 6.12 |
| $Q_{da,left}$ [ml/s] | 1.00 ± 0.18 | 1.00 ± 0.26 |
| $Q_{dm,left}$ [ml/s] | 3.75 ± 0.45 | 3.72 ± 0.72 |
| $Q_{dp,left}$ [ml/s] | 1.50 ± 0.22 | 1.49 ± 0.31 |
| $P_c$ [mmHg] | 25.03 ± 2.03 | 24.95 ± 3.38 |
| $Q_{pv}$ [ml/s] | 12.49 ± 1.66 | 12.41 ± 2.36 |
| $P_v$ [mmHg] | 14.04 ± 0.62 | 14.02 ± 1.41 |
| $Q_{out}$ [ml/s] | 12.50 ± 0.66 | 12.42 ± 1.61 |

**Table 1. Mean and standard deviation values of computed cerebrovascular variables stratified by SR and AF simulations.** $P_a$: systemic arterial pressure; $Q_{ICA,left}$: left internal carotid flow rate; $Q_{BA}$: basilar artery flow rate; $P_{BA,willis}$: basilar artery pressure at the entrance of the Circle of Willis; $Q_{ACA1,left}$: left precommunicating anterior cerebral artery flow rate; $Q_{PCA1,left}$: left precommunicating posterior cerebral artery flow rate; $P_{ACA,left}$: left anterior cerebral artery pressure; $P_{MCA,left}$: left middle cerebral artery pressure; $P_{ACA,left}$: left posterior cerebral artery pressure; $Q_{ACA2,left}$: left postcommunicating anterior cerebral artery flow rate; $Q_{PCA2,left}$: left postcommunicating posterior cerebral artery flow rate; $Q_{MCA,left}$: left middle cerebral artery flow rate; $P_{da,left}$: left anterior distal pressure; $P_{dm,left}$: left middle distal pressure; $P_{dp,left}$: left posterior distal pressure; $Q_{da,left}$: left anterior distal flow rate; $Q_{dm,left}$: left middle distal flow rate; $Q_{dp,left}$: left posterior distal flow rate; $P_c$: cerebral capillary pressure; $Q_{pv}$: proximal venous flow rate; $P_v$: cerebral venous pressure; $Q_{out}$: total outflow rate.

To assess whether the greater variability of hemodynamic parameters during AF may relate to critical hemodynamic events, we computed the number of hypoperfusions and hypertensive events. Figure 6 reports the absolute frequency, evaluated over 5,000 cardiac cycles, of hypoperfusions and hypertensive events along the different districts of the ICA-MCA pathway during AF (according to the adopted definition, neither hypoperfusions nor hypertensive events occurred during SR simulation). In particular, a total number of 303 hypoperfusions (maximum duration: 2 beats) occurred at the distal arterial circulation ($Q_{dm,left}$), while 387 hypertensive events (maximum duration: 5 beats) occurred at the capillary level ($P_c$).

## Discussion

Several mechanisms (SCI, microbleeds, altered cerebral blood flow dynamics, pro-inflammatory conditions) potentially play a role in the genesis of AF-associated cognitive impairment/dementia[3,4]. Cerebral blood flow dynamics during AF is the least explored field: available data are mostly dated[17,18] and seem to indicate that AF is associated with reduced mean cerebral blood flow, at least in patients under 65 years of age. More recently, AF patients with rapid ventricular rates treated by atrioventricular node ablation and pacemaker implantation were observed to improve brain perfusion and cognitive function[19], through a systolic and cardiac output increase, only partially hinting a relation between atrioventricular synchronization and cerebral perfusion. However, arteriolar and capillary hemodynamics in terms of time-variable pressure and flow rate signals during AF have, to the best of our knowledge, never been investigated.

In this scenario, the present lumped-parameter model of the cerebrovascular circulation, taking into account cerebral autoregulation, is a valuable tool to investigate and compare AF and SR cerebral hemodynamics. In this computational study, AF and SR are considered in absence of other associated pathologies, such as hypertension, that could themselves affect cerebral hemodynamics: as a consequence, the reported results strictly focus on the hemodynamic impact exerted by heart rhythm alone on the cerebral circulation.

In general, flow rate outcomes in the SR model are within the physiologic range. In the present model mean total cerebral blood flow is 12.5 ml/s, while in a study by Zarrinkoob et al.[20], over 49 healthy young patients, the value averaged 12.85 ml/s. Afferent blood flow partitions are also similar: $Q_{ICA,left}$: 4.75 ml/s vs. 4.6 ml/s and $Q_{BA}$: 3 ml/s vs. 2.7 ml/s in the present model and the study of Zarrinkoob et al.[20], respectively. In addition, the partition downstream of the Circle of Willis is also comparable, with a predominant contribution of MCA with respect to ACA and PCA, confirming that the present model provides a realistic evaluation of the cerebral hemodynamics.

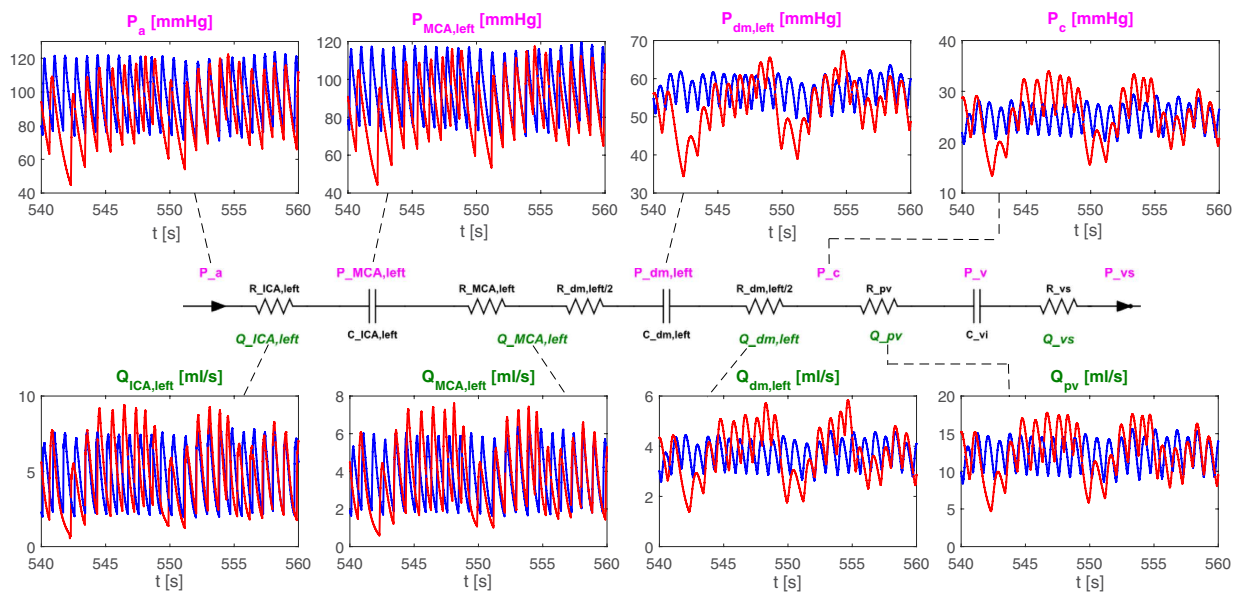

**Figure 4. Pathway ICA-MCA.** Representative pressure and flow rate series are reported for the selected pathway during SR (blue) and AF (red).

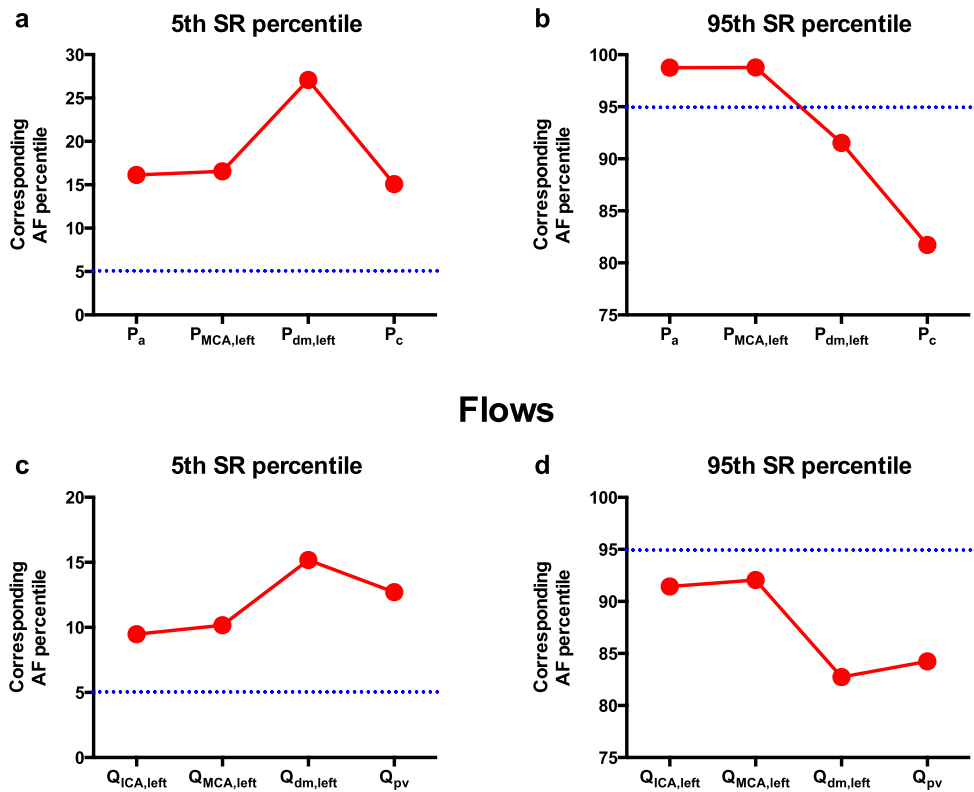

**Figure 5. Percentile evaluation.** Evaluation of AF percentiles corresponding to the SR thresholds (5% SR and 95% SR) along the ICA-MCA pathway: (**a**) Flow rate, 5% SR; (**b**) Flow rate, 95% SR; (**c**) Pressure, 5% SR; (**d**) Pressure, 95% SR.

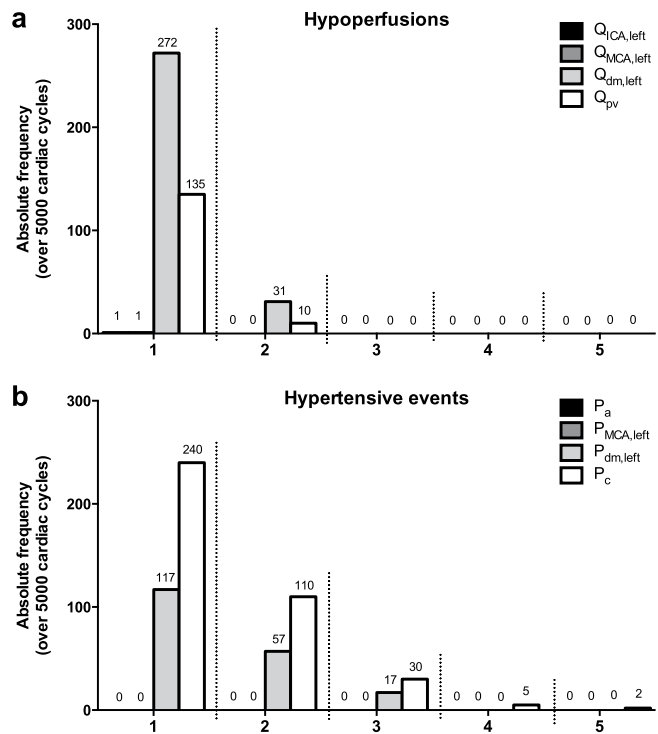

**Figure 6. Absolute frequency of hypoperfusions and hypertensive events during AF along the ICA-MCA pathway.** (**a**) Hypoperfusions; (**b**) hypertensive events. The abscissa indicates the number of consecutive beats characterizing the events.

Based on the computed cerebrovascular variables (Table 1), it clearly emerges that, even though pressures are generally slightly reduced during AF (due to reduced systemic pressure), mean flow rates in the vessels are similar between SR and AF models. This finding highlights that a well-functioning cerebral autoregulation system is able to ensure a normal mean cerebral flow during AF. Instead, the major difference between SR and AF simulations is apparent in the standard deviations of each cerebrovascular variable. During AF, standard deviations present higher variability compared to SR simulations: in particular, the variability of the cerebral cardiovascular parameters tends to increase in the distal circulation, reaching maximum variability at the arteriolar and capillary levels (Fig. 5).

In these districts (arterioles and capillaries) AF also alters the pulsatile dynamics of the cerebrovascular variables, as clearly depicted in Fig. 4, resulting in local hypoperfusions and hypertensive events, lasting from a single to several consecutive beats. Concerning hypoperfusions (Fig. 6a), they are more prevalent at the arteriolar level (303 events), with the longest events lasting 2 consecutive beats, while hypertensive events (Fig. 6b) tend to occur mostly at the capillary level (387 events) and to last longer (up to 5 consecutive beats). The shorter duration of the hypoperfusions, compared to hypertensive events, may reflect autoregulation mechanism's priorities, which more stringently force to regain critical mean cerebral flow values.

According to the present computational findings, the hemodynamic cerebral effect of AF can be a relevant mechanism into the genesis of AF-related cognitive impairment/dementia. In fact, deep white matter could undergo an ischemic damage either due to the transient hypoperfusions or as a consequence of being exposed to transient hypertensive events (by arteriolosclerosis and capillary loss/bleeding), laying the basis for a potential AF-related vascular subcortical dementia[21,22]. Therefore, the hemodynamic alterations demonstrated in our model may explain at least a subgroup of non-embolic AF-related SCI and non anticoagulant-related cerebral microbleeds.

**Limitations.** The following limitations should be taken into account. First, the computational model, both for AF than SR, does not consider the impact that rate control drugs (e.g. digoxin, beta blockers, non-dihydropiridine calcium channel blockers) could exert on the cardiovascular system. Second, the cerebrovascular model assumes a perfectly functioning cerebral autoregulation mechanism in both SR and AF. Lastly, the cardiovascular model used to obtain the forcing input, $P_a$, does not account for short-term pressure regulation effects of the baroreceptor mechanism. In addition, from a mechanistic point of view, sinus rhythm with frequent ectopy (>1,000/24 hours)[23] could lead to isolated patterns which partially resemble those observed throughout AF rhythm.

## Conclusion

Several mechanisms have been proposed to explain the AF-related increased risk of dementia and cognitive decline, independent of clinically relevant events (e.g. strokes and TIAs). Concerning cerebral blood flow dynamics, the presented computational approach suggests that the major difference between AF and RS stands in the higher variability of the hemodynamic variables during the former. In particular, this effect tends to increase in the distal circulation, reaching a maximum at the arteriolar and capillary levels, resulting in local transient periods of excessive pressure or reduced blood flow.

Based on the findings demonstrated in our model, the impact of AF on cerebral hemodynamics in the distal circulation, *per se,* is as a potential mechanism related to the genesis of AF-related cognitive impairment/dementia.

## Author Contributions

## Additional Information

**Supplementary information** accompanies this paper at http://www.nature.com/srep

**Competing financial interests:** The authors declare no competing financial interests.

**How to cite this article**: Anselmino, M. *et al.* Transient cerebral hypoperfusion and hypertensive events during atrial fibrillation: a plausible mechanism for cognitive impairment. *Sci. Rep.* **6,** 28635; doi: 10.1038/srep28635 (2016).